\documentclass[pla,twocolumn,superscriptaddress,showpacs]{revtex4}
\usepackage{amssymb,graphicx}

\begin{document}

\title{Curvature-induced quantum behaviour on a helical nanotube}

\author{Victor Atanasov}
 \altaffiliation[Also at]{ Laboratoire de Physique Th\'{e}orique et
Mod\'{e}lisation , Universit\'{e} de Cergy-Pontoise,
 F-95302 Cergy-Pontoise, France }
\affiliation{Institute for Nuclear Research and Nuclear Energy,
Bulgarian Academy of Sciences,  72 Tsarigradsko chaussee, 1784
Sofia, Bulgaria}
 \email{victor@inrne.bas.bg}

\author{Rossen Dandoloff}
\affiliation{ Laboratoire de Physique Th\'{e}orique et
Mod\'{e}lisation , Universit\'{e} de Cergy-Pontoise,
 F-95302 Cergy-Pontoise, France}
 \email{rossen.dandoloff@u-cergy.fr}

\begin{abstract}
We investigate the effect of curvature on the behaviour of a
quantum particle bound to move on a surface shaped as a helical
tube. We derive and discuss the governing Schr\"odinger equation and the corresponding quantum effective potential which is periodic and points to the helical configuration as more energetically favorable as compared to the straight tube. The exhibited periodicity also leads to energy band structure of pure geometrical origin. 
\end{abstract}


\maketitle

Recent developments in nanotechnology\cite{AFS} made it possible
to grow quasi-two-dimensional surfaces of arbitrary shape where
quantum and curvature effects play a major role\cite{MT}. Examples
include single crystal $\rm NbSe_3$ M\"obius strips\cite{TTOIYH},
spherical $\rm CdSe-ZnS$ core-shell quantum dots\cite{SWFEB},
$\rm Si$ nanowire, nanoribbon transistors\cite{DNSPEG}, quantum
waveguides\cite{LCM} and nanotorus\cite{torus}. Several
publications\cite{INTT,Clark*96,Exner*95,Exner*01,Fujita*04,Kaplan,Mitchell} have
treated the constrainment of quantum-mechanical particles
(with applications in, e.g. standard Sch\"odinger equation
problems\cite{Jaffe*03} and relativistic Dirac equation
problems\cite{Jaffe*99,BJ*93}) to a two-dimensional surface since the
original works by Jensen and Koppe, da
Costa\cite{JK*71,daCosta*81,daCosta*82}. Since two-dimensional systems are an a priori idealization it is reasonable to quantize before constraining the particle to the nanotube. As a result a quantum particle confined to a
two-dimensional surface embedded in $\mathbb{R}^3$ experiences a
potential that is a function of the Mean and the Gauss curvatures of
the surface\cite{daCosta*81,daCosta*82}. This curvature-induced quantum potential is a geometrical
invariant which property lead the authors\cite{vic*07} to pose
the inverse differential geometrical problem: what curved surfaces
produce prescribed curvature-induced potential.

Possible physical applications of the above include the geometric interaction between defects and curvature in thin
layers of superfluids, superconductors, and liquid crystals
deposited on curved surfaces\cite{Vit*04}; the
curvature of a semiconductor surface determines also an
interesting mechanism of spin--orbit interaction of
electrons\cite{Ent*01}; a charged quantum particle
trapped in a potential of quantum nature due to bending of an
elastically deformable thin tube travels without dissipation like
a soliton\cite{Dan*05} ; the twist of a strip plays a role of a
magnetic field and is responsible for the appearance of localized
states and an effective transverse electric field thus reminisce
the quantum Hall effect\cite{Dan*04}.

Now let us turn our attention to the geometrical realization of the helical tube.
\begin{figure}[b]
\begin{center}
\includegraphics[scale=0.35]{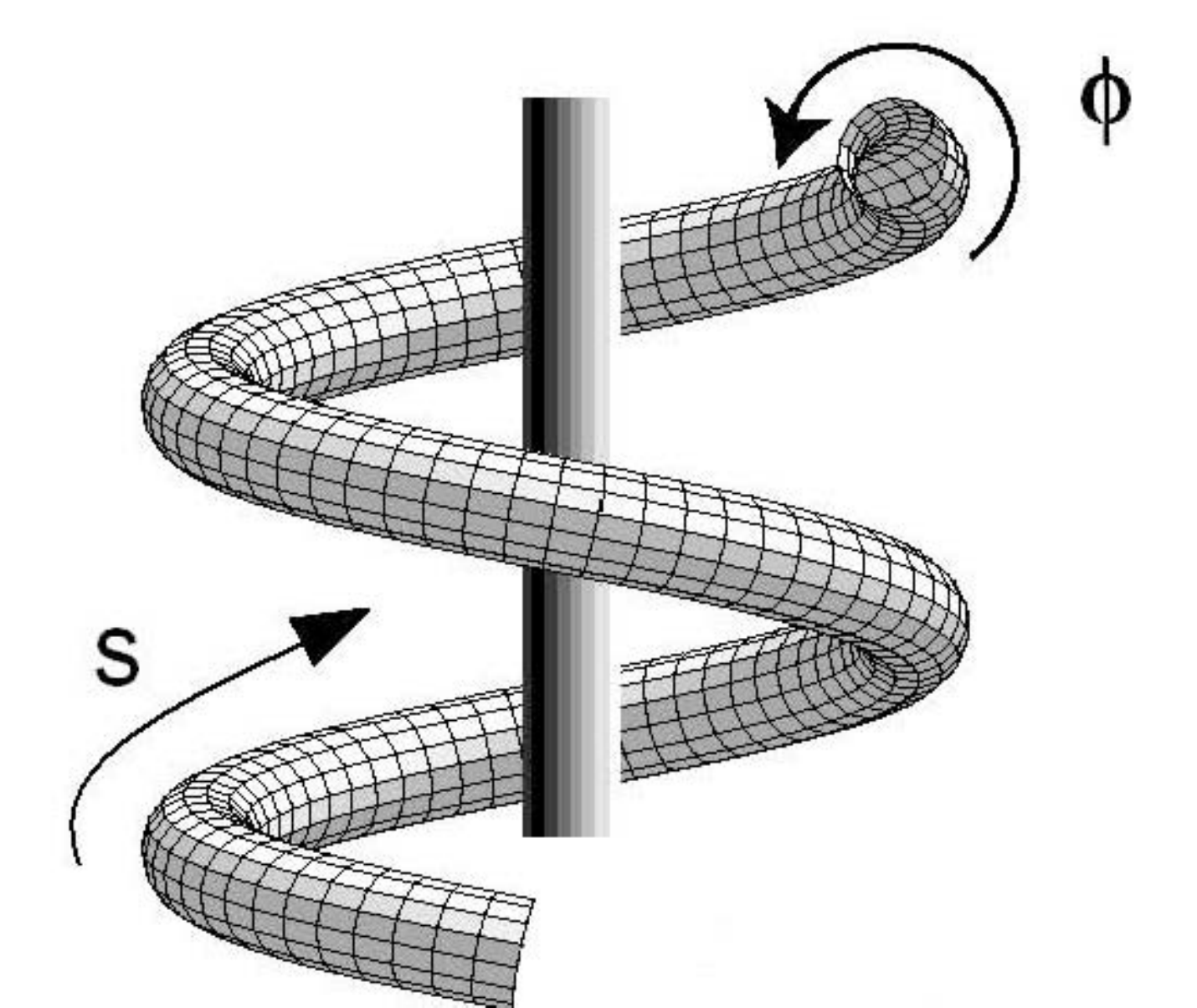}
\caption{\label{fig:tube} The geometry of an infinite helical tube
may be parametrized by two families of space curves (see equation
(\ref{eq:coord_X}) and text).}
\end{center}
\end{figure}
One can associate with a  space curve $\vec{\bf x}(s)$ at any
point $s$ along it a moving frame consisting of three
vectors $\vec{\bf t}$--tangent, $\vec{\bf n}$--normal and
$\vec{\bf b}$--binormal and evolving along the curve according to
the Frenet-Serret equations:
\begin{eqnarray}\label{eq:Frenet}
\dot{\vec{{\bf t}}}=\vec{\omega}
\wedge \vec{{\bf t}},\qquad \dot{ \vec{{\bf b}}}=\vec{\omega} \wedge \vec{{\bf
b}},\qquad \dot{
\vec{{\bf n}}}=\vec{\omega} \wedge \vec{{\bf n}},
\end{eqnarray}
where $\vec{\omega}$ is the instantaneous angular velocity of the
Frenet-Serret frame where the arclength $s$ plays the role of time. Hereafter the dot denotes derivation with
respect to the natural parameter $s.$ Here $\kappa(s)$
and $\tau(s)$ are the curvature and torsion of the space curve.

Since $\vec{\omega}$ has a component along $\vec{{\bf t}}$ we
redefine the frame vectors
\begin{eqnarray}\label{eq:N}
\vec{{\bf N}}=\cos{\theta(s)}\vec{{\bf n}}+
\sin{\theta(s)}\vec{{\bf b}},&& \dot{ \vec{{\bf N}}}=\vec{\Omega}
\wedge \vec{{\bf N}}  \\\label{eq:B} \vec{{\bf
B}}=-\sin{\theta(s)}\vec{{\bf n}}+ \cos{\theta(s)}\vec{{\bf b}},&&
\dot{\vec{{\bf B}}}=\vec{\Omega} \wedge \vec{{\bf B}}.
\end{eqnarray}
We choose $\theta(s)$ so that $\vec{\Omega}$ has no component in
the direction of $\vec{{\bf t}}$. A brief calculation yields
\begin{equation}\label{eq:theta}
\theta(s)=-\int_{s_0}^{s} d s^{'} \tau(s^{'}).
\end{equation}

\begin{figure}[t]
\begin{center}
\includegraphics[scale=0.70]{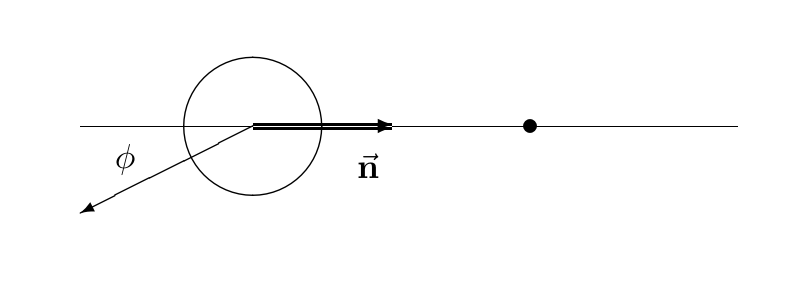}
\caption{\label{fig:cross-section} The cross-section of the nanotube in Fig. \ref{fig:tube}. }
\end{center}
\end{figure}

Now let us mount a disc $D$ rigidly in the reference frame where
$\vec{{\bf N}}$ and $\vec{{\bf B}}$ are at rest, i.e. the
Fermi-Walker frame\cite{Gold&Jaffe*92}. The points on the surface may be
parametrized by
\begin{equation}\label{eq:coord_X}
\vec{{\bf X}}(s, \phi)=\vec{\bf x}(s) - \rho_0\left\{\sin{\phi}
\vec{{\bf B}}+\cos{\phi} \vec{{\bf N}}\right\}.
\end{equation}

The two families of space
curves weaving the above surface in $\mathbb{R}^3$ are the following.
The first is a circle parametrized by the
angle $\phi$ and is actually the rim of the disc that is rigidly
mounted to the tangent of $\vec{\bf x}(s)$ at each point in space. The tip of the vector in the disc from the central axis to the rim is denoted by $\rho_0\cos{\phi} \vec{{\bf N}}+\rho_0\sin{\phi}
\vec{{\bf B}}.$ Its origin coincides with the helical space line
$\vec{\bf x}(s).$ The second is given by the lines with tangent passing
through each point of the first family. Refer to Figure \ref{fig:tube} for the visual expression of the above construction.

In this article we will study the properties of the
Schr\"odinger equation on that surface.

The line element is
\begin{equation}\label{eq:g}
|d \vec{{\bf X}}|^2=d \varphi^2 + h^2 d s^2,
\end{equation}
where
\begin{equation}\label{eq:h}
h(s, \phi)=1+\rho_0
\kappa(s)\cos{\left[\theta(s)+\frac{\varphi}{\rho_0}\right]}
\end{equation}
and
\begin{equation}
\varphi=\rho_0 \phi
\end{equation}
has a dimension of length.

If we change the
parametrization $s \rightarrow -s$ and $\varphi \rightarrow
-\varphi$ this would mean that we evolve the surface backward from a certain
arbitrary point $s_0$ of the infinite space line $\vec{\bf x}(s).$ The torsion $\tau$
exhibits invariance $\tau(s)=\tau(-s)$ and the surface element must remain unchanged. 
\[
\theta(-s)+\left(-\frac{\varphi}{\rho_0}\right) \rightarrow
-\left[\theta(s)+\frac{\varphi}{\rho_0}\right], \quad h(s)
\rightarrow h(-s).
\]
Thus we show that the line element is indeed invariant
\[
|d \vec{{\bf X}}(s,\varphi)|^2=|d \vec{{\bf X}}(-s,-\varphi)|^2.
\]

From formulas (\ref{eq:N}) and (\ref{eq:B}) we see that at
$\theta(s)=0$, that is at $s=0$ if $s_0=0$ (see (\ref{eq:theta})),
we have the coincidence $\vec{{\bf N}} \equiv \vec{{\bf n}}$ and
$\vec{{\bf B}} \equiv \vec{{\bf b}}.$ The normal $\vec{{\bf n}}$
always points towards the axis around which the helix is wound,
i.e. it points inward. From (\ref{eq:h}) it is clear that
$h(0,0)=1+\rho_0 \kappa(0) > 1 - \rho_0 \kappa(0)=h(0,\pi).$ The
surface is stretched more on the outside thus we have a natural
choice of the origin (the outer intersection of the ray through
$\vec{{\bf n}}$ and the cross-section of the tube) for the two families of
curves (see Figure \ref{fig:cross-section}).

Introducing the normal to the surface
$\vec{\nu}$  from the Gauss triad
$
\vec{\nu} = \partial_{\varphi} \vec{{\bf X}} \wedge
\partial_{s} \vec{{\bf X}} \left| \partial_{\varphi} \vec{{\bf X}} \wedge
\partial_{s} \vec{{\bf X}}\right|^{-1/2}
$
we can compute the linear Weingarten map
$
\left(%
\begin{array}{c}
  \partial_{\varphi} \vec{\nu} \\
  \partial_s \vec{\nu} \\
\end{array}%
\right)= W
\left(%
\begin{array}{c}
  \partial_{\varphi} \vec{{\bf X}} \\
  \partial_s \vec{{\bf X}} \\
\end{array}%
\right),
$
where $W$ is the matrix realizing the map of the tangent space in itself
\begin{equation}\label{eq:W}
W=\left(%
\begin{array}{cc}
  \rho_0^{-1} & 0 \\
  0 & \kappa(s)\cos\left[\theta(s)+\frac{\varphi}{\rho_0}\right]h^{-1} \\
\end{array}%
\right).
\end{equation}
With the help of (\ref{eq:W}) we may compute
\[
M=\frac12 (\kappa_1+\kappa_2)=-\frac12{\rm tr}(W), \quad K=\kappa_1 \kappa_2=\det(W),
\]
the Mean and the Gauss curvatures of the surface respectively,
where $\kappa_1$ and $\kappa_2$ are the principal curvatures of
the surface. They are also the eigenvalues of the
Weingarten matrix (\ref{eq:W}). Thus we obtain
\begin{equation}\label{eq:k1}
\kappa_1=\frac{1}{\rho_0}, \qquad
\kappa_2 = \kappa(s)
\cos\left[\theta(s)+\frac{\varphi}{\rho_0}\right]  h^{-1}.
\end{equation}

\begin{figure}[t]
\begin{center}
\includegraphics[scale=0.47]{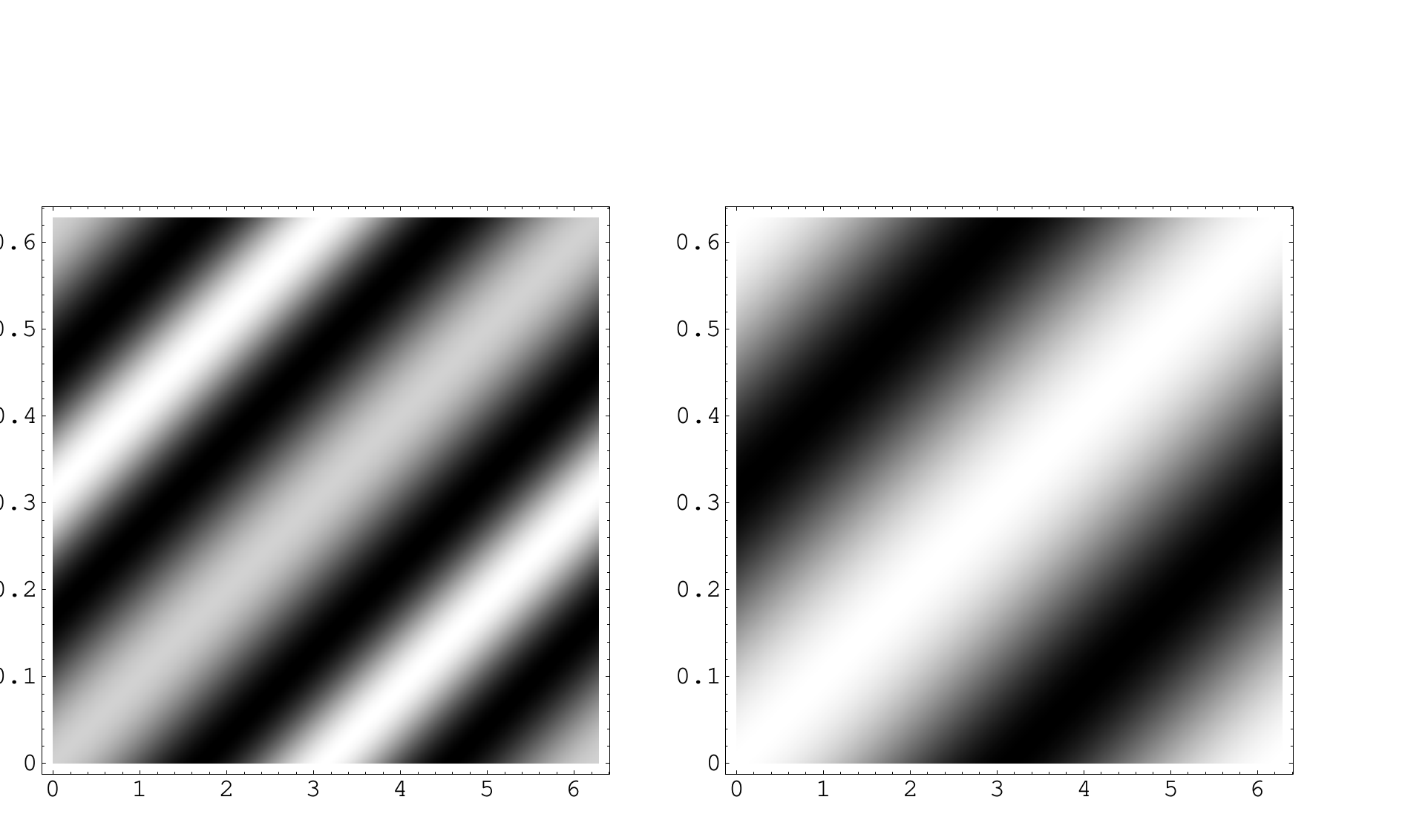}
\caption{\label{fig:pot} Density plot of the potentials $V_{\rm
eff}(s,\varphi)$ on the left and $V_{\rm curv}(s,\varphi)$ on the right.
Along the horizontal axis is the natural parameter $s$,
where $s \in [0,2\pi]$; $\phi$ is on the vertical axis
$\varphi \in [0,\rho_02 \pi]$. Here $\tau=\kappa=1$ and
$\rho_0=10^{-1}$. Lighter regions correspond to higher values of
the potential and lower probability to find a particle there,
respectively.}
\end{center}
\end{figure}

Since we study the resulting Schr\"odinger equation for a particle confined to
move on that surface and following da Costa an
effective potential appears in the Schr\"odinger  equation which
has the following form:
\begin{eqnarray}
V_{\rm
curv}&=&-\frac{\hbar^2}{2 \mu}(M^2-K)\\
\nonumber &=&-\frac{\hbar^2}{2\mu}\left[\frac14
({\rm tr}W)^2-\det(W)\right]
\end{eqnarray}
where $\mu$ is the effective particle's mass,
$\hbar$--Plank's constant; $V_{\rm curv}$ depends on $s$ and
$\varphi$ which appear as the generalized coordinates on the surface;
$M=(\kappa_1+\kappa_2)/2$ and $K=\kappa_1 \kappa_2$ are the Mean
and the Gauss curvatures respectively. For the surface (\ref{eq:coord_X})  we obtain

\begin{equation}\label{eq:Vcurv}
V_{\rm curv}(s,\varphi)=-\frac{\hbar^2}{8\mu} \frac{1}{\rho_0^2} \frac{1}{h^2}.
\end{equation}

From equations (\ref{eq:g}) and (\ref{eq:h}) it follows that the surface is more stretched on the outside, that is at $\varphi=0$ (see Figure \ref{fig:cross-section}), because $h(0,0)>h(0,\pi).$
The Heisenberg uncertainty principle states that a particle would have a lower energy where the line element is bigger. Our expectation is that the
probability to find a particle on the outer rim of the surface is
maximal. This guiding principle will allow us to interpret the appropriate effective Schr\"odinger equation whose potential possesses the above property.

The Laplace-Beltrami operator (the quantum mechanical kinetic
term) in the coordinate system (\ref{eq:coord_X}) can be written as follows:
\begin{eqnarray}\label{eq:LaplacePsi}
\nonumber - \triangle_{s,\varphi} \Psi &=&- \frac{1}{h^{2}} \frac{\partial^2
\Psi}{\partial s^2} - \frac{\partial^2 \Psi}{\partial \varphi^2} +
 \kappa \sin{\left(\theta+
\frac{\varphi}{\rho_0}\right)} \frac{1}{h}
\frac{\partial \Psi}{\partial \varphi} \\
&+& \rho_0 \dot{\kappa}(s)
\cos{\left(\theta(s)+\frac{\varphi}{\rho_0}\right)}\frac{1}{h^{3}}
\frac{\partial \Psi}{\partial s}\\
\nonumber &-& \rho_0 \dot{\theta}(s)\kappa(s)
\sin{\left(\theta(s)+\frac{\varphi}{\rho_0}\right)}
\frac{1}{h^{3}}
\frac{\partial \Psi}{\partial s}.
\end{eqnarray}
Here $\Psi$ as a solution must be normalized as $\int |\Psi|^2 dS=1.$
We introduce $\Phi=\sqrt{h} \Psi$ and the wave function will be normalized with respect to the usual flat norm on a rectangular domain determined by the periodic properties of $h[\theta(s), \varphi],$ that is  $\int_{-\rho_0 \pi}^{\rho_0 \pi}  d \varphi \int_{s_a}^{s_b} d s |\Phi|^2  =1,$ where $s_a$ and $s_b$ are such that $\theta(s_a)=-\theta(s_b)=\pi.$ Then:

\begin{eqnarray}\label{eq:LaplacePhi}
- \sqrt{h} \triangle_{}\frac{ \Phi}{\sqrt{h}} &=& - \frac{1}{h^{2}} \frac{\partial^2
\Phi}{\partial s^2} - \frac{\partial^2 \Phi}{\partial \varphi^2}  + 2
\frac{\partial_s h}{h^{3}}
\frac{\partial \Phi}{\partial s}  + V_{\rm kin} \Phi, \qquad
\end{eqnarray}
where

\begin{eqnarray}\label{eq:Vkin}
V_{\rm kin}= \frac12 \frac{\partial_{\varphi}^2 h}{h^{}} - \frac14 \frac{(\partial_{\varphi} h)^2}{h^{2}}
 + \frac12 \frac{\partial_{s}^2 h}{h^{3}} - \frac54 \frac{(\partial_{s} h)^2}{h^{4}}.
\end{eqnarray}

Now we obtain a differential equation
which for the helical tube is to be written with $\kappa$ and
$\tau$ constants:
\begin{eqnarray}\label{eq:SchrodingerPhi}
- \frac{1}{h^{2}} \frac{\partial^2
\Phi}{\partial s^2} - \frac{\partial^2 \Phi}{\partial \varphi^2} +2
\frac{\partial_s h}{h^{3}}
\frac{\partial \Phi}{\partial s}  + V_{\rm eff} \Phi - k^2 \Phi=0,
\end{eqnarray}
where
\begin{eqnarray}\label{eq:Veff}
 V_{\rm eff} (s,\varphi)&=& V_{\rm kin} +\frac{2 \mu} {\hbar^2}  V_{\rm curv} \\
\nonumber &=&  \frac{\partial_{\varphi}^2 h}{2h^{}} -   \frac{(\partial_{\varphi} h)^2}{4h^{2}}
 + \frac{\partial_{s}^2 h}{2h^{3}} - \frac54 \frac{(\partial_{s} h)^2}{h^{4}} -\frac14 \frac{1}{\rho_0^2 h^2}.
\end{eqnarray}

Here $k$ is the
wave number ($k^2=2\mu E/\hbar^2$). The corresponding spatial configuration for which equation (\ref{eq:SchrodingerPhi}) serves as an effective Schr\"odinger equation is depicted in Figure \ref{fig:tube}.

Fixing $s=0$ a straight-farward check provides us with the estimate
\[
V_{\rm eff}(\varphi=0) <  V_{\rm eff}(\varphi=\pi)
\]
and since the probability amplitude follows the behaviour of potential the probability to find the particle on the outer rim of the surface is greater in accordance with the Heienberg's principle. Figure \ref{fig:pot}  presents the density plots of the two potentials. 

Now we will expand the curvature induced effective potential
$V_{\rm eff}$ and the kinetic operator in series.
Next we assume $\kappa/\tau \sim 1$
and since we want to acquire insight into the properties of
nanosystems we set $\rho_0 \kappa = \epsilon \ll 1$ ($\rho_0$ represents the radius of the nanotube, measured in nanometers) thus  a small parameter naturally arises. Now we expand in series the denominators up to first
order terms in
$\epsilon$ and equation (\ref{eq:SchrodingerPhi}) reduces to an effective two dimensional perturbed Schr\"odinger equation on a rectangular domain in $\mathbb{R}^2:$

\begin{eqnarray}\label{eq:SchrodingerPhiExpanded}
\frac{\partial^2
\Phi}{\partial s^2} + \frac{\partial^2 \Phi}{\partial \varphi^2}  + k_{\rm eff}^2 \Phi=V^{(1)} \Phi ,
\end{eqnarray}
where the perturbing potential is
\begin{eqnarray}\label{eq:V1}
\nonumber V^{(1)}& =& \epsilon \left\{  \frac12 \kappa^2 \left[   \cos{\left(\tau s-
\frac{\varphi}{\rho_0}\right)}  +  \cos^2{\left(\tau s-
\frac{\varphi}{\rho_0}\right)} \right. \right. \\
& -& \left. \left. \cos^3{\left(\tau s-
\frac{\varphi}{\rho_0}\right)}       \right] +  \cos{\left(\tau s-
\frac{\varphi}{\rho_0}\right)} \partial_{s}^2   \right.  \\
\nonumber &-& \left.   \tau  \sin{\left(\tau s-
\frac{\varphi}{\rho_0}\right)}  \partial_{s}    \right\}
\end{eqnarray}
and
\begin{eqnarray}\label{eq:kEFF}
k_{\rm eff}^2 &=&  a + \mathcal{E},\quad  a=\frac14 \left( \frac{1}{\rho_0^2}  + \kappa^2 \right), \quad \mathcal{E}= \frac{2 \mu E}{\hbar^2}.\quad
\end{eqnarray}
Notice, that due to the presence of the squared curvature {\it the helical configuration is more energetically favorable as compared to the straight tube} where this term vanishes. This may favor the helical shape in the experimentally grown nanotubes\cite{MAZAMN}.

It would be interesting to elaborate on the consequences of the limit $\epsilon=0.$ In this case the geometry goes to that of a cylinder and we expect to recover the corresponding results\cite{Willatzen}. Indeed the dependence of the wave function $\Phi$ on $s$ is of the form of a standing wave in a one--dimensional box stretching to infinity with the boundary conditions $\Phi(s=0)=\Phi(s=L)=0$ as $L \to \infty.$ The corresponding eigenenergies are vanishing
\[
E_{\rm l} = \frac{\hbar^2}{2 \mu} \left( \frac{ {\rm l} \pi}{L} \right)^2, \qquad L \to \infty.
\]
What we are left with is a harmonic oscillator equation for the dependence of $\Phi$ on $\varphi.$ The required periodicity of the solution introduces a new quantum number ${\rm n}$ in (\ref{eq:kEFF}) which leeds to
\[
E_{\rm n,l}=\frac{\hbar^2}{2 \mu \rho_0^2}\left[ {\rm n}^2- \frac14 + \rho_0^2\left( \frac{{\rm l} \pi}{L} \right)^2 \right], \qquad L \to \infty.
\]
This is in exact agreement with the results on the finite cylinder\cite{Willatzen}.

Due to the periodicity of the coefficients both in $\varphi$ and $s$ the wave function must be periodic and we may look for a solution as a Bloch wave function\cite{Madelung}, that is
\begin{equation}\label{eq:Bloch}
\Phi=\frac{1}{\sqrt{S}} e^{i \vec{k} \vec{r}} \sum_m u(\vec{K}_m) e^{i \vec{K}_m \vec{r}},
\end{equation}
where $S=(2 \pi)^2 \rho_0/\tau$ is the area of the two-dimensional  rectangular domain determined by the symmetry of the problem and the periodicity of the wave function. In the above
\begin{equation}
\vec{r}=\left( \begin{array}{c}
 s\\
 \varphi \\
\end{array} \right), \qquad  \vec{K}_m= \left( \begin{array}{c}
 m_s \tau \\
 m_\varphi \rho_0^{-1} \\
\end{array} \right),
\end{equation}
where $m_s$ and $m_{\varphi}$ are integers and
\begin{equation}\label{eq:u(0)}
u(0) \sim 1.
\end{equation}
What is assumed  small in this approximation is the perturbing potential $V^{(1)}$ of the order of $\epsilon.$ It also contains derivations which for the reasonably well-behaved wave function (\ref{eq:Bloch}) produce no singularities and the order of smallness is preserved.

Inserting this ansatz into (\ref{eq:SchrodingerPhiExpanded}) we obtain
\begin{eqnarray}\label{eq:FourierSchrodinger}
[   - (  \vec{k} +  \vec{K}_n )^2 &+& k^{2}_{\rm eff} ] u(\vec{K}_{n}) \\
\nonumber &=&    \tilde{V}^{(1)} (\vec{K}_n -\vec{K}_m)     u(\vec{K}_m),
\end{eqnarray}
where $\tilde{V}^{(1)} $ are the coefficients of the expansion of the perturbing potential (\ref{eq:V1}) along the inverse lattice
\begin{equation}
 V^{(1)} (\vec{r}) =  \sum_{p  } \tilde{V}^{(1)} (\vec{K}_m)    e^{i \vec{K}_m \vec{r}}.
\end{equation}

Within  the approximation (\ref{eq:u(0)}) we obtain
\begin{equation}\label{eq:u(K_n)}
u(\vec{K}_m) =  \frac{ \tilde{V}^{(1)} (\vec{K}_m ) }{ k^2_{\rm eff}  - \left(  \vec{k} +  \vec{K}_m  \right)^2} .
\end{equation}
Thus the coefficients in the expansion of the Bloch wave function (\ref{eq:Bloch}) have the same order of magnitude as ${V}^{(1)}$ and it suffices to add only few of them in order to obtain the behaviour of the wave function on the surface of the helical tube.

It can easily be seen that due to the argument of the periodic functions in (\ref{eq:V1}) we can only have non-vanishing contribution along one ray in the inverse lattice, e.g. the ray associated with the vector
\begin{equation}
\vec{K}_1= \left( \begin{array}{c}
 \tau \\
 - \rho_0^{-1} \\
\end{array} \right), \qquad \vec{K}_{-1}= - \vec{K}_{1}.
\end{equation}
Thus the direction in the inverse lattice where the sum in (\ref{eq:Bloch}) is performed is determined.

The exhibited singularity at
 $k^2_{\rm eff} = \left(  \vec{k} +  \vec{K}_m  \right)^2$
 in (\ref{eq:u(K_n)}) suggests that not only $u(0)$ but also $u(\vec{K}_m )$ component may be considered as ''big enough''. We may write a system of equations for the two components directly from (\ref{eq:FourierSchrodinger}) which gives
\begin{eqnarray}
\nonumber && \left[   - \left(  \vec{k} +  \vec{K}_m   \right)^2 + k^2_{\rm eff} \right] u(\vec{K}_m) - \tilde{V}^{(1)} (\vec{K}_m ) u(0) =0, \\
 && -  \tilde{V}^{(1)} (-\vec{K}_m)     u(\vec{K}_m) + \left[   - \vec{k}^2 + k^2_{\rm eff} \right] u(0) =0.
 \end{eqnarray}
Equating to zero the determinant of the above system is necessary condition for solvability and produces the expressions for the energies due to (\ref{eq:kEFF}) in adjacent zones. Introducing the notation $U^2(\vec{K}_m)=\tilde{V}^{(1)} (\vec{K}_m)   \tilde{V}^{(1)} (-\vec{K}_m)$ we have
\begin{widetext}
\begin{eqnarray}\label{eq:E12}
\mathcal{E}_{1,2}& =& \frac12 \left[ 2 a  - \vec{k}^2 -\left(  \vec{k} +  \vec{K}_m   \right)^2 \right]     \pm  \sqrt{ \frac14 \left[ 2 a  - \vec{k}^2 -\left(  \vec{k} +  \vec{K}_m   \right)^2 \right]^2  -  \left(a  - \vec{k}^2 \right) \left[ a  -\left(  \vec{k} +  \vec{K}_m   \right)^2 \right]  + U^2    },
\end{eqnarray}
\end{widetext}
where each root describes an energy band. Here $a$ is given by (\ref{eq:kEFF}) and encodes the curvature dependence. It is convenient to expand the energy in terms of $\vec{G}$ which measures the difference $\vec{G} = \vec{k} + \frac12 \vec{K}_m$ in wave vector between $\vec{k}$ and the zone boundary at $-\frac12 \vec{K}_m.$ In the region where $\vec{K}_m^2 \vec{G}^2 \ll U^2 $ 

\begin{eqnarray}
\mathcal{E}_{1,2}& =& \mathcal{E}(\pm)  -\left( \frac12 \vec{K}_m   \right)^2  \left[ 1   \mp  2 \frac{  \vec{G}^2}{\left| U \right|}    \right],
\end{eqnarray}
where $\mathcal{E}(\pm)=a  - \vec{G}^2 \pm \left| U \right|,$ so the energy has 2 roots, one lower than the free electron kinetic energy $a  - \vec{G}^2$(shifted due to the presence of curvature in $a$) by $\left| U \right|$ and one higher by $\left| U \right|.$ Thus the curvature-induced potential has created an energy gap $2 \left| U \right|$ at the zone boundary. The above expression is valid only when the wave vector is very close to the zone boundary. The gap in the energy spectrum opened in the transition between the first and the second zones in the inverse lattice scales approximately as 
\begin{equation}
2 \left| U \right| \sim \epsilon \left( \frac{\kappa^2}{4} \right),
\end{equation}
where ${\kappa^2}/{4}$ is the curvature-induced potential due to da Costa of a space line whose curvature is $\kappa.$ It is of pure geometrical origin. In this case the space line is helix and the energy gap scales as the ratio between the diameter of the nanotube and its radius of curvature, that is $\epsilon,$ times the energy due to the curved configuration.

Let us note that the effective mass tensor  
$
M_{ij}=\left( \frac{1}{\hbar^2} \frac{\partial^2 E(\vec{k})}{\partial k_i \partial k_j} \right)^{-1} \neq \mu \delta_{ij}
$
is diagonal. Here the particle acquires an effective mass due to the interaction with curvature. This interaction is encoded in the presence of the square root containing $a$ in (\ref{eq:E12}) which remains after twice differentiating with respect to the wave vector components in accordance with the conveyed formula for the mass.

In conclusion for a helical nanotube we have obtained an effective Schr\"odinger equation which is periodic. The quantum effective potential shows that it is more probable to find a quantum particle on the outer rim of the nanotube. As a whole the helical configuration is more energetically favorable as compared to the straight tube. The properties of the effective Schr\"odinger equation are discussed within the Bloch ansatz and a gap in the energy spectrum is shown to arise in the transition between adjacent zones of the inverse lattice. It is geometry dependent and can be tested also experimentally.

\end{document}